\documentclass{desyproc}

\begin{document}
\title{The Hunt for neutrinoless double beta decay with the NEXT experiment}

\author{{\slshape David Lorca$^1$ on behalf of the NEXT Collaboration}\\[1ex]
$^1$Instituto de Fisica Corpuscular (IFIC), CSIC \& Univ. de Valencia, E-46071 Valencia, Spain}

\contribID{65}

\confID{65}  
\desyproc{DESY-PROC-2014-04}
\acronym{PANIC14} 
\doi  

\maketitle

\begin{abstract}
The NEXT-100 detector will search for the neutrinoless double beta decay of $^{136}$Xe using an electroluminescent high-pressure xenon gas TPC filled with 100~kg of enriched Xe. An observation of this hypothetical process would establish a Majorana nature for the neutrino and prove the violation of lepton number.  A scaled-down prototype, NEXT-DEMO, has been built to demonstrate the feasibility of the technology. NEXT-DEMO includes an energy plane made of PMTs and a tracking plane made of SiPMs. X-ray energy depositions, produced by the de-excitation of xenon atoms after their interaction with gamma rays, have been used to characterize the detector response. With this method, the released energy by gammas coming from $^{22}$Na source has been corrected, achieving an energy resolution of 5.691\%~FWHM and 1.62\%~FWHM at the 29.7~keV and 511~keV peaks respectively, which extrapolate to 0.62\%~FWHM and 0.73\%~FWHM at Q$_{\beta \beta}$ value of Xenon.
\end{abstract}

\section{Introduction}

Double beta decay ($\beta \beta$) is a very rare nuclear transition in which a nucleus with Z protons decays into a nucleus with Z+2 protons and same mass number A. It can only be observed in those isotopes where the $\beta$ decay mode is forbidden due to the energy of the daughter nuclei being higher than the energy of the parent nuclei, or highly suppressed. Two decay modes are usually considered: the standard two-neutrino mode ($\beta \beta 2 \nu$), which has been observed in several isotopes with typical half-lives in the range of $10^{18} - 10^{21}$ years~\cite{GomezCadenas:2011it}, and the neutrinoless mode ($\beta \beta 0 \nu$), which violates lepton-number conservation, and is therefore forbidden in the Standard Model of particle physics. 

An observation of $\beta \beta 0 \nu$ would prove that neutrinos are Majorana particles, that is, identical to their antiparticles~\cite{Schechter:1981bd}, and would provide direct information on neutrino masses~\cite{GomezCadenas:2011it}. Besides, it would demonstrate that total lepton number is violated, a result that can be linked to the cosmic asymmetry between matter and antimatter through the process known as leptogenesis~\cite{Davidson:2008bu}.

The half-life of $\beta \beta 0 \nu$, if mediated by light, Majorana neutrino exchange, can be written as 
\begin{equation}
(T^{0\nu} _{1/2})^{-1} = G^{0\nu}  \left |M^{0\nu}\right |^2 m^2 _{\beta \beta}
\end{equation}
where $G^{0\nu}$ is an exactly-calculable phase-space integral for the emission of two electrons; $\left |M^{0\nu}\right |$ is the nuclear matrix element of the transition, which has to be evaluated theoretically; and $m _{\beta \beta}$ is the effective Majorana mass of the electron neutrino:
\begin{equation}
m _{\beta \beta} = \left | \sum U^2 _{ei}  m_i\right |
\end{equation}
where $m_i$ are the neutrino mass eigenstates and $U_{ei}$ are elements of the neutrino mixing matrix.

The aim of all $\beta \beta 0 \nu$ experiments is to measure the decay rate of this disintegration. However, the measurement is limited by the experimental sensitivity of the detector employed, which can be expressed as
\begin{equation}
T_{1/2} \propto a \cdot{} \epsilon \cdot{} \sqrt{\frac{M\cdot{} t}{\Delta E \cdot {B}}}
\end{equation}
where $M$ is the isotope mass, $\Delta E$ is the energy resolution, $B$ is the background rate, $\epsilon$ is the detection efficiency and $a$ is a term which includes nuclear matrix elements~\cite{GomezCadenas:2011it}.

Due to the presence of the two neutrino mode, together with background events which can fall in the energy Region of Interest (ROI) where the neutrinoless mode is expected, $\Delta E$ is a must to resolve the possible $\beta \beta 0 \nu$ events. In addition, an appropriate selection of detector components and surroundings should be done in such a way to reduce background rate as low as possible. Besides, current generation of $\beta \beta 0 \nu$ experiments have explored the region of neutrino masses corresponding to 160-250~meV (depending of n.m.e.)~\cite{Gando:2012zm} by using from tens to a few hundred kilos of isotope mass. The non detection of a signal creates the necessity of increase the isotope mass in the new generation up to the ton scale to explore new areas.

\section{The NEXT Concept}

The NEXT experiment combines good energy resolution, a low background rate and the possibility to scale-up the detector to large masses of $\beta \beta$ isotope by using a high-pressure xenon gas (HPXe) electroluminescent (EL) time projection chamber (TPC) to search for $\beta \beta 0 \nu$ in $^{136}$Xe. 

Whit this technology, an energy resolution better than 1\%~FWHM can be achieved in NEXT at Q$_{\beta \beta}$ of Xe thanks to the small Fano Factor of gaseous xenon ($F_{HPXe} = 0.15\pm0.02$)~\cite{Nygren2009337}, compared with other media such as liquid xenon (LXe) ($F_{LXe} \sim 20$), together with the low fluctuations introduced by an EL-based amplification. Besides, HPXe provides topological information of the events, allowing to discriminate between signal events (a twisted track of about 10~cm long, with two energy depositions at both ends) from background events (single electrons with only one blob at the end and most of the time accompanied by an X-ray~\cite{Collaboration:2012ha}). Furthermore, $^{136}$Xe constitutes 8.86\% of all natural xenon, but the enrichment process is relatively simple and cheap compared to that of other $\beta \beta$ isotopes, thus making $^{136}$Xe the most obvious candidate for a future multi-ton experiment.

The detection process in NEXT implies independent systems for tracking and calorimetry. Particles interacting in the HPXe transfer their energy to the medium ionizing and exciting its atoms. The excitation energy is manifested in the prompt emission of VUV $(\sim178\ \rm{nm})$ scintillation light. The ionization electrons drift toward the TPC anode thanks to the presence of a moderate electric field, entering in a region with an even more intense electric field. There, secondary VUV photons are generated isotropically by electroluminescence. Therefore both scintillation and ionization produce an optical signal, to be detected with a plane of PMTs (the energy plane) located behind a transparent cathode. The detection of the primary scintillation light constitutes the start-of-event, whereas the detection of EL light provides an energy measurement. Electroluminescent light provides tracking as well, since it is detected a few millimeters away from production at the anode plane, via an array of MPPCs (the tracking plane).

\section{NEXT-DEMO: R\&D and results}

To demonstrate that the NEXT concept is feasible, a scaled prototype, NEXT-DEMO, was developed. NEXT-DEMO is a cylindrical pressure vessel made of stainless steel, able to withstand up to 20~bar of internal pressure. It is 60~cm long and 30~cm in diameter, and holds $\sim$~1.5~kg of Xe at 10~bar. Three wire grids, the cathode, gate and anode, limit the two active regions of the TPC. The primary scintillation light is directly detected by a plane of 19 Hamamatsu R7378A PMTs behind the cathode grid. Electroluminescent light produced by ionization electrons, is once again detected in the energy plane but the forward going photons are also detected in an array of 256 tetraphenyl butadiene (TPB) coated Hamamatsu S10362-11-050P SiPMs. The tracking plane is used to reconstruct the position of energy deposits and, ultimately, the topology of an event as a whole.

In this prototype, the abundance of xenon K-shell X-ray emission during data taking with a $^{22}$Na source has been identified as a multitool for the characterisation of the fundamental parameters of the gas as well as the equalisation of the response of the detector~\cite{Lorca}. The advantage of using these events is that they are distributed all over the volume of the detector and the range of the $\sim$30~keV electrons produced is small, around 0.6~mm at 10 bar~\cite{ESTAR}, releasing almost all their energy in a single point. Such depositions have been used to extract correction factors which describe the detector geometry effects. In addition, both loss of charge due to electron attachment with gas impurities and temporal fluctuations during the EL generation due to temperature and pressure oscillations have been corrected.

The mentioned corrections have been applied to the reconstructed energy released by gammas coming from a $^{22}$Na (see~\cite{Lorca}), where an energy resolution for the K$_\alpha$ peak (29.7~keV) and photopeak (511~keV) of (5.691 $\pm$ 0.003)\% FWHM and (1.62 $\pm$ 0.01)\% FWHM were extracted respectively. Independently extrapolating these two values to the $^{136}$Xe $Q_{\beta \beta}$ assuming the dominance of photon shot noise Poisson statistics results in a predicted energy resolution at $Q_{\beta \beta}$ of 0.6256\% FWHM  and 0.7353\% FWHM respectively.

\section{NEXT-100}

Following the previous ideas, the NEXT collaboration plans to build the NEXT-100 detector, described in~\cite{Collaboration:2012ha}, which will be formed by a HPXe TPC containing 100~kg of xenon, enriched at 90\% in its $^{136}$Xe isotope, at 15~bar. The pressure vessel is built with low activity stainless steel, and contains an inner copper shield, 12~cm thick and made of radio pure copper, to attenuate the radiation coming from the high-energy gammas emitted in the decays of $^{208}$Tl and $^{214}$Bi, present in the external detector. The energy measurement in NEXT-100 is provided by a total of 60 Hamamatsu R11410-10 photomultipliers (PMTs) covering 32.5\% of the cathode area constitute the energy plane. This PMT model has been specially developed for radiopure, xenon-based detectors. The tracking function is provided by an array of around 7200 SiPMs, 1~cm pitch, located behind the EL region, and coated with TPB.

All materials present in the NEXT-100 detector have been chosen according to rigorous radiopurity requirements, which together with the detection technique employed by NEXT, produce an expected background rate of 5$\cdot{}$10$^{4}$ counts/(kg$\cdot{}$keV$\cdot{}$year)~\cite{Collaboration:2012ha}. After 5 years of data taking, a sensitivity of 5.9$\cdot{}$10$^{25}$ years is predicted or, in terms of the effective neutrino Majorana mass m$_{\beta \beta}$, a value of around 100 meV, making NEXT one of the most competitive experiments in the field~\cite{GomezCadenas:2010gs}.

NEXT-100 is approved for operation in the Laboratorio Subterr\'aneo de Canfranc (LSC), in Spain, where the installation of seismic platform, lead castle, gas system, emergency recovery system and vessel are already completed. Underground operations with non-enriched xenon will start in 2015 and the physics case with enriched xenon is planned for early 2016.

\section{Conclusions}

The search for $\beta \beta 0 \nu$ is one of the major current challenges in neutrino physics. Due to the high sensitivity provided by a HPXe TPC with EL amplification, NEXT-100 promises to be one of the leading experiments in the field, exploring the region of neutrino mass down to 100~meV. One of its prototypes, NEXT-DEMO, has demonstrated the main issues of such technology, helping in the design of the final detector. In addition, xenon K-shell X-ray depositions have been identified as a useful tool for the characterization of this type of detectors, providing the spatial calibration needed for close-to-optimal energy resolution.

\section{Acknowledgments}

This work was supported by the following agencies and institutions: the European Research Council under the Advanced Grant 339787-NEXT; the Ministerio de Econom\'ia y Competitividad of Spain under grants CONSOLIDER-Ingenio 2010 CSD2008-0037 (CUP), FPA2009-13697-C04 and FIS2012-37947-C04; the Director, Office of Science, Office of Basic Energy Sciences, of the US Department of Energy under contract no.\ DE-AC02-05CH11231; and the Portuguese FCT and FEDER through the program COMPETE, project PTDC/FIS/103860/2008. 



\begin{footnotesize}


\end{footnotesize}


\end{document}